\documentclass[twocolumn]{aastex62}
\received{January 15, 2020}
\accepted{February 15, 2020}

\shorttitle{Two Component Jets of GRB160623A as Shocked Jet cocoon afterglow}
\shortauthors{Chen Urata et al.}

\begin{document}

\title{Two Component Jets of GRB160623A as Shocked Jet cocoon afterglow}

\correspondingauthor{Yuji Urata}
\email{urata@g.ncu.edu.tw}

\author{Wei Ju Chen}
\affiliation{Institute of Astronomy, National Central University, Chung-Li 32054, Taiwan}

\author{Yuji Urata}
\affiliation{Institute of Astronomy, National Central University, Chung-Li 32054, Taiwan}

\author{Kuiyun Huang}
\affiliation{Center for General Education, Chung Yuan Christian University, Taoyuan 32023, Taiwan}

\author{Satoko Takahashi}
\affiliation{Joint ALMA Observatory, Alonso de Cordova 3108, Vitacura, Santiago, Chile}
\affiliation{NAOJ Chile Observatory, Alonso de Córdova 3788, Oficina 61B, Vitacura, Santiago, Chile}
\affiliation{Department of Astronomical Science, School of Physical Sciences, SOKENDAI (The Graduate University for Advanced Studies), Mitaka, Tokyo 181-8588, Japan}

\author{Glen Petitpas}
\affiliation{Harvard-Smithsonian Center for Astrophysics, 60 Garden Street, Cambridge, Massachusetts 02138, USA}

\author{Keiichi Asada}
\affiliation{Academia Sinica Institute of Astronomy and Astrophysics, Taipei 106, Taiwan}


\begin{abstract}

Two components of jets associated with the afterglow of the gamma-ray
burst GRB 160623A were observed with multi-frequency observations
including long-term monitoring in a sub-millimetre range (230 GHz)
using the SMA.  The observed light curves with temporal breaks
suggests on the basis of the standard forward-shock synchrotron
radiation model that the X-ray radiation is narrowly collimated with
an opening angle $\theta_{n,j}<\sim6^{\circ}$ whereas the radio
radiation originated from wider jets ($\sim27^{\circ}$).  The temporal
and spectral evolutions of the radio afterglow agree with those
expected from a synchrotron radiation modelling with typical physical
parameters except for the fact that the observed wide jet opening
angle for the radio emission is significantly larger than the
theoretical maximum opening angle.  By contrast, the opening angle of
the X-ray afterglow is consistent with the typical value of GRB jets.
Since the theory of the relativistic cocoon afterglow emission is
similar to that of a regular afterglow with an opening angle of
$\sim30^{\circ}$, the observed radio emission can be interpreted as
the shocked jet cocoon emission.  This result therefore indicates that
the two components of the jets observed in the GRB 160623A afterglow
is caused by the jet and the shocked jet cocoon afterglows.

\end{abstract}
%


\section{Introduction} \label{sec:intro}

The Gamma-ray burst (GRB) is believed to be a stellar explosion
accompanied with relativistic outflows and narrowly-collimated
jets \citep[e.g.][]{piran99}. Since direct imaging of GRB jets is impossible
unlike AGN jets, the jet opening angles of GRBs have been measured by
identifying a temporal break in the light curve in multi-frequency
afterglow monitoring \citep{sari99}.  The typical value of GRB jet
opening-angles is $\sim3^{\circ}.5$ \citep[e.g.][]{racusin09}, which is in the
same order with that of AGN jets (median of $1^{\circ}.5$ among 373
samples) measured with high-resolution imaging observations
\citep{agn}.
For both the two populations of the GRBs, short and long GRBs,
understanding of the jet and its structure is essential.
There are several methods to constrain the GRB jet structure.  An
optical spectroscopic study of an associated supernova component has
identified a cocoon structure \citep{izzo19}.  Another method is to
measure the detection ratio of off-axis GRB afterglows without prompt
high-energy emissions \citep[i.e. orphan GRB afterglow e.g.][]{N02}. However,
systematic detection of orphan GRB afterglows has never been made \citep[e.g.][]{orphan}.
Continuous multi-frequency afterglow monitoring is another crucial
method to constrain the jet structure. In the case of GRB030329, 
double-component jets (narrow and wide jets)  were identified  with 
optical and radio monitoring, including sub-millimetre
\citep{grb030329}.
%
\begin{figure*}[htb]
\centering
\includegraphics[width=1.8\columnwidth]{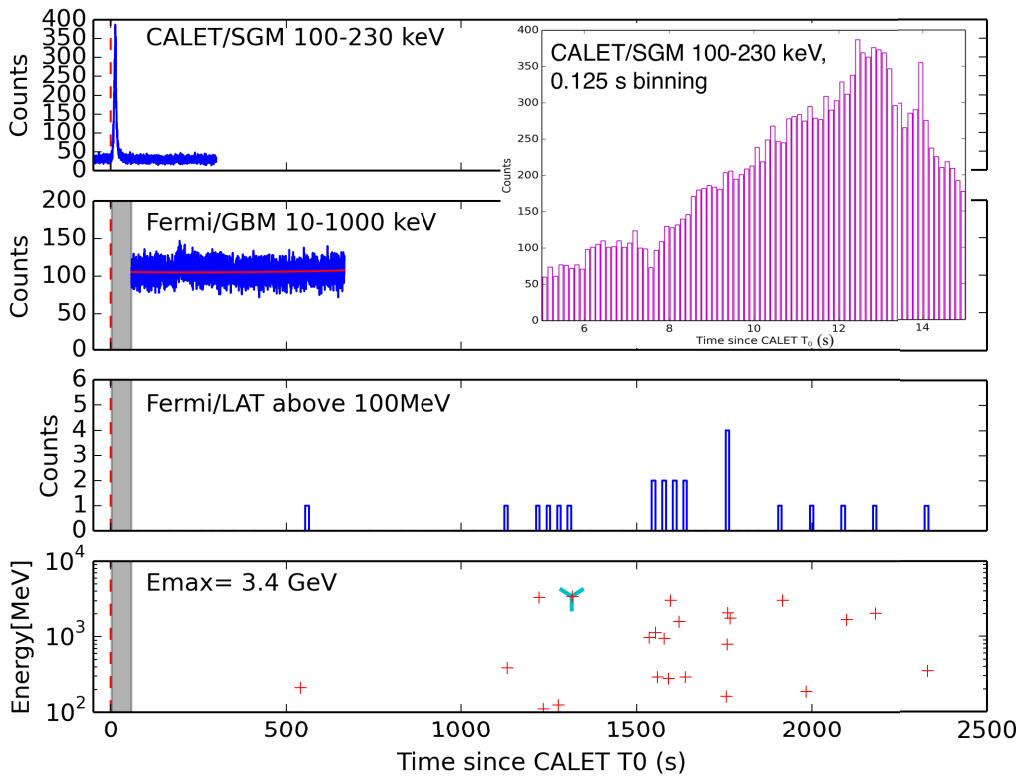}
\caption{ Light curves of (top 3 panels) counts of GRB 160623A in the prompt phase observed  with CALET, Fermi/GBM, and Fermi/LAT  and of (bottom panel) the photon energy distribution observed  with Fermi/LAT. The grey shaded parts indicate the interval unobserved with {\it Fermi} due to  Earth occultation. The inset shows a zoomed-up time-series  at around the main peak in the top panel. Variability on a timescale of as short as 0.250 sec is visible.
}
\label{prompt-lc}
\end{figure*}
%
Sub-millimetre and millimetre afterglow observations have  played
an essential role in revealing  new insights of the GRB afterglow
\citep[e.g.][]{urata14,huang17,urata19}. 
Here, we report the long-term
monitoring of the GRB160623A afterglow using the Sub-millimeter Array (SMA)
 in conjunction with multi-frequency observations.
We characterise the dependence of the afterglow flux on time and frequency
as $F(t,\nu)\propto t^{\alpha}\nu^{\beta}$, where $\alpha$ is the decay
index and $\beta$ is the spectral energy index.  We use the
cosmological parameters of $\Omega_{M}=0.3$, $\Omega_{\Lambda}$=0.7,
and $H_{0}=70$ km s$^{-1}$ Mpc$^{-1}$ in this paper.

\section{Observations and Results} \label{sec:obs}

\begin{figure*}[htb]
\centering
\includegraphics[width=1.9\columnwidth]{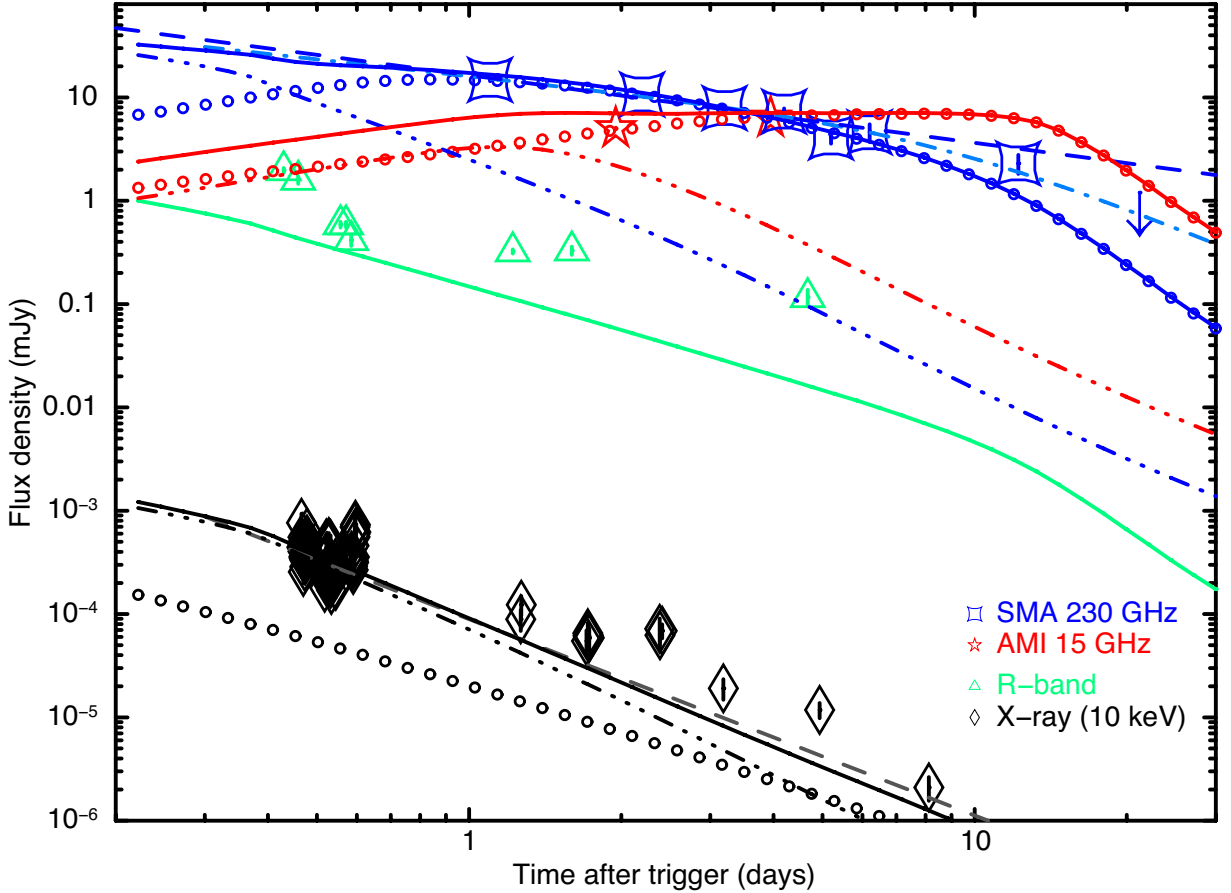}
\caption{ Light curves of the GRB 160623A afterglow in (black) X-rays, (green) optical R-band, (blue) 230 GHz, and (red) 15 GHz. Blue dash-dotted line indicates the best-fitting smooth broken power-law function of the 230-GHz light curve. The dashed lines show simple power-law functions with (black) $\alpha_{X}=-1.92$ and (blue) $\alpha_{230GHz}=-0.65$. The open-circle-dotted lines in (black) X-rays, (green) optical, (blue) 230 GHz, and (red) 15 GHz show the best-fitting synchrotron radiation models for the radio data (i.e. 15 and 230 GHz). The dash-dot-dot-dot lines in (black) X-rays, (blue) 230 GHz, and (red 15 GHz) show the expected narrow jet component based on the collapsar jet case. The solid lines show the total model function (i.e. summention of narrow and wide components). 
}
\label{lc-all}
\end{figure*}

\subsection{Prompt emission}

The Fermi Gamma-Ray Monitor (GBM) found a signal triggered by GRB 160623A at 05:00:34.23
UT on 2016 June 23 \citep{gcn-gbm}. The Fermi Large Area Telescope (LAT)
also detected more than 15 photons above 1 GeV  till approximately 2
ksec (0.02 days) after the trigger time and  determined the centre position at (RA,
Dec) = (315.24, +42.27$^\circ$) (J2000) with an error radius of
0.1$^\circ$ \citep{gcn-lat}.
GRB160623A was  detected also by CALET Gamma-ray Burst Monitor (CGBM)
at 04:59:34.27 on 2016 June 23,  which was 1 min earlier than the Fermi GBM
trigger time \citep{gcn-calet}. Hereafter, we use the trigger time of
the CGBM as the burst starting time, ${\rm T}_{0}$. All of the CGBM
instruments detected the emission and the light curves exhibited a
bright peak at ${\rm T}_{0}$+40 sec.  By contrast, two of the Fermi instruments
(GBM and LAT) missed to observe the main peak of the event.
The Konus-Wind was also triggered at 04:59:37.594 and detected the
emission up to $\sim$15 MeV \citep{gcn-konus}.  The time-averaged
spectrum for the main burst in the 10 keV--10 MeV range
 was described by  a Band function with  low and high-energy photon  indices of
$\alpha=-0.76^{+0.02}_{-0.02}$ of $\beta =
-2.80^{+0.05}_{-0.06}$, respectively, and a peak energy $E^{obs}_{p} =
596^{+15}_{-14}$ keV. The equivalent isotropic radiated energy in the prompt phase at the 10 keV--10MeV band $E_{\rm iso}$ was estimated as $(2.53\pm0.03)\times10^{53}$ erg \citep{gcn-konus}.

We obtained the light-curve data with 0.125-sec time bins observed with
the CALET from the CGBM Flight Trigger Alert Notices
site\footnote{http://cgbm.calet.jp/cgbm\_trigger/flight/}.  We
measured 5$\sigma$ flux variations relative to the neighbouring data
bins
for a timescale of 0.250 sec (Figure \ref{prompt-lc}).

The Fermi/GBM data  were downloaded from the NASA HEASARC Fermi GBM Burst
catalog.  We used the Fermitools version 1.0.7 and HEASOFT for reducing
the {\it Fermi} GBM/LAT data with $gtsrcprob$ p $>$ 0.9 and a GTI selection of
``DATA\_QUAL$>$0, 
LAT\_CONFIG$==$1, 
and 
ABS(ROCK\_ANGLE)$<$52''.  The user contribution code  ``\texttt{do{\_}gbm.py}'' by
S. Holland was used for the GBM light-curve analysis.
The {\it Fermi}/LAT photon data were downloaded from the Fermi Science
Support Center.   Using the likelihood and aperture photometry, we
generated the light curve for  an energy range of $>$100 MeV  (Figure
\ref{prompt-lc}). The highest-energy photon within the 2500 sec time
coverage was $\sim$3.4 GeV at 1315 sec after the burst,  which was
considerably  after the main pulse observed  with CALET and
Konus-Wind. According to the Fermi/LAT GRB catalog \citep{latgrbcat},
the energetic photon  at 18 GeV was also observed at 12038.53 sec
after the burst.

Figure \ref{prompt-lc} shows the light curves obtained by CALET/SGM
(100--230 keV), Fermi/GBM (10--1000 keV), and Fermi/LAT ($>$ 100 MeV)
 along with the photon energy distribution for an energy higher than 100 MeV.

\begin{figure}[htb]
\centering
\includegraphics[width=0.95\columnwidth]{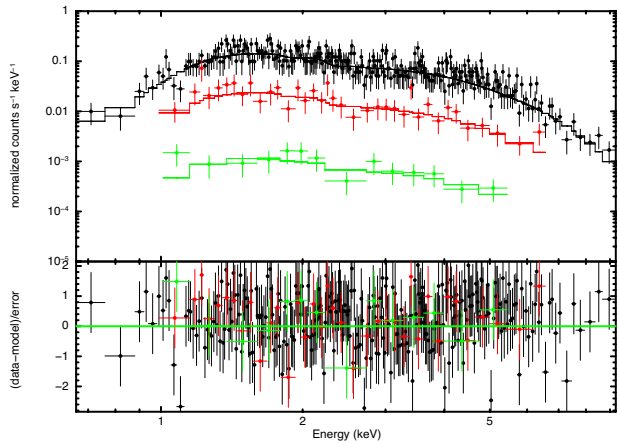}
\caption{X-ray afterglow spectra  in (black) 0.47--0.60 days, (red) 1.3--2.4 days, (green) 3.2--11.5 along with the best-fitting model with an absorbed power law. Bottom panel shows the residual  of the fitting. For the first  period (black), we set the intrinsic absorption as a free parameter. The derived best-fitting values of the absorption column density and spectral index were N$_{\rm H}$=$(2.7\pm0.3)\times10^{22}$ cm$^{-2}$ and $\beta_{X}=-0.92\pm0.10$, respectively. For the later  periods, we fixed the intrinsic absorption obtained with the first  period of spectrum and derived the spectral index  to be $\beta_{X}=-1.0\pm0.18$ and $\beta_{X}=-0.89\pm0.33$, respectively.
}
\label{xspec}
\end{figure}
\begin{table*}
\begin{center}
\label{tab:smaflux}
\title{}
\caption{SMA observations}
\begin{tabular}{lllllll}
\hline \hline
Date  & t$_{start}$ &  t$_{end}$ & N$_{antennas}$ &  time from trigger & flux density & flux error \\
 & & & &(days with mid of observation) &(mJy) & (mJy) \\
\hline 
2016/06/24 & 07:49:04.4 & 15:19:45.4 & 5 &  1.274 & 14.8 & 0.5 \\                  
2016/06/25 & 10:10:03.5 & 15:58:49.8 & 7 &  2.336 & 10.7 & 0.4 \\                   
2016/06/26 & 10:33:10.9 & 16:02:09.9 & 7 &  3.345 & 7.9 & 0.4 \\                   
2016/06/27 & 10:18:05.6 & 18:05:56.5 & 7 &  4.383 & 7.3 & 0.5 \\
2016/06/28 & 10:12:26.2 & 18:00:26.1 & 7 &  5.379 & 4.1 & 0.5 \\
2016/06/29 & 10:24:08.2 & 17:45:43.3 & 7 &  6.378 & 4.6 & 0.9 \\
2016/07/05 & 10:19:21.5 & 17:14:25.3 & 7 & 12.365 & 2.3 & 0.5 \\
2016/07/14 & 10:23:49.6 & 16:42:46.6 & 8 & 21.356 & 1.2 & (3-$\sigma$ upper limit) \\
\hline
\end{tabular}
\end{center}
\end{table*}

\subsection{Afterglow}
\subsubsection{X-ray and optical follow-ups}

Neil Gehrels Swift Observatory started follow-up observations at
$\sim40$ ksec after the burst. The XRT identified the X-ray afterglow at
R.A. = 21$^{h}$01$^{m}$11$^{s}$.22, Decl.=
+42$^{\circ}$13$^{'}$13$^{"}$.7 with an  error radius of
3".5 \citep{swiftxrt}. The X-ray afterglow was observed  with the XRT
until $\sim$12 days after the burst.  The UVOT also obtained images with
$u$ and $v$ bands and no counterparts in the bands  were
observed \citep{uvot}.


The optical afterglow was  detected and its position was determined  to be R.A. =
21$^{h}$01$^{m}$11$^{s}$.65, Decl.= +42$^{\circ}$13$^{'}$15$^{"}$.0
 with Nordic Optical Telescope \citep[NOT;][]{gcn-not}.  The
photometric observations of the optical afterglow were executed  with 
Murikabushi \citep{murikabushi}, RATIR \citep{ratir}, NOT,
AZT-33IK \citep{azt} and Zeiss-1000 \citep{zeiss}.  
The redshift was determined to be $z=0.367$ from the H$_\alpha$, S II and N II
emission lines obtained with  the NOT and the Gran Telescope Canarias \citep[GTC;][]{gtc}.

We obtained reduced light curves and spectra in three  periods of 0.47--0.60 days,
1.3--2.4 days, and 3.2--11.5 days  of the {\it Swift}/XRT data
from the  UK Swift Science Data Centre \citep{xrt1,xrt2}.  
%
The X-ray light curve  is found to be described with  a single power-law function
with  a decay index of $\alpha_{X}=-1.92\pm0.04$  with a reduced
$\chi^2$/dof=1.04/86 (Figure \ref{lc-all}).  
%
%
We rebinned the spectra so that each spectral bin contains more than 5
counts.  Using the software XSPEC 12, we perform spectral fitting with
a single power law modified with intrinsic and Galactic absorptions,
the latter of which is fixed at N$_{\rm H}=7.17\times10^{21}$
cm$^{-2}$.
For the first period, we perform spectral fitting, allowing the
intrinsic absorption column density to vary.  The derived best-fitting
values of the intrinsic absorption column density and spectral index
are N$_{\rm H}$=$(2.7\pm0.3)\times10^{22}$ cm$^{-2}$ and
$\beta_{X}=-0.92\pm0.10$, respectively, with a reduced
$\chi^2$/dof=0.88/276 (Figure \ref{xspec}). For the later periods, we
fix the intrinsic absorption to the value obtained with the first
period of spectrum.  The derived spectral indices are
$\beta_{X}=-1.0\pm0.18$ for the second period with a reduced
$\chi^2$/dof=0.71/35 and $\beta_{X}=-0.89\pm0.33$ for third period
with a reduced $\chi^2$/dof=0.63/12.
We therefore find no spectral evolution after 0.47 days, comparing the spectra
at three periods of 0.47--0.60 days, 1.3--2.4 days, 3.2--11.5 days.

\subsubsection{Submillimeter Array and radio follow-ups}

We executed sub-millimetre (230 GHz) follow-up observations using the
SMA. The first continuum observation was performed on 2016 June 24,
about 1.1 days after the burst.  The observation identified a bright
($\sim15$ mJy) submm afterglow, which is one of the brightest GRB
afterglows ever detected in the submillimetre range \citep{submmgrb2}.
Continuous monitoring was then performed at the same frequency
setting on 2016 June 25, 26, 27, 28, 29, and July 5 and 14 (Table 1).
We reduced the SMA data, using the MIR data-reduction package and
Miriad software. 
The data were flagged and calibrated with the MIR data-reduction
package, using the standard procedure, and then images were
constructed, using the Miriad software. The total flux was measured
with the Common Astronomy Software Applications \citep[CASA;][]{32}.

We fit the SMA light-curve with a simple power-law function.  The
fitting using the time range from 1.3 to 12.4 days (i.e. all
detections) yields a power-law index $\alpha=-0.65\pm0.07$ with a
reduced $\chi^2$/dof=4.4/5. Note that the fitting would be
significantly improved if we selected the period before 5 days. The
temporal decay is described by the simple power-law with
$\alpha=-0.54\pm0.05$ (reduced $\chi^{2}$/dof = 1.3/2).  In addition,
the extrapolation of the above-mentioned steeper index
(i.e. $\alpha=-0.65$) is inconsistent with the upper limit of 21.4
day.
Hence, these results indicate that there is a gradual temporal break
after $\sim$12 days.
We employ a smoothly-connected broken power-law function with a
smoothness parameter of 1 and fixed decay indices before and after the
break as $-0.54$ and $-2$, respectively. The fitting yields the
temporal break at $t_{R,j}=27\pm14$ days.

The AMI Large Array detected the radio afterglow at 15 GHz and
measured the brightness to be $5.0\pm0.1$ mJy at 2.0 days and
$6.3\pm0.1$ mJy at 4.0 days, respectively \citep{ami}. These
measurements indicate that the light curve at 15 GHz exhibited a
brightening with $\alpha\sim0.33$ between 2.0 and 4.0 days.
The radio spectral indices between the AMI and SMA bands are also
found to be $\beta=\sim0.27$ at 2 days and $\beta=\sim0.05$ at 4 days.

\section{Discussion} \label{sec:discuss}
\subsection{Radiation of Afterglow}

The closure relation \citep[e.g. summarized in][]{2004IJMPA..19.2385Z} indicates that the X-ray
afterglow after 0.46 days was consistent with the relation
$\nu_{c}<\nu_{X}$ during the post jet-break phase with the index of
the electron energy distribution, $p<2$ (i.e., $\alpha =(\beta-3)/2$).
The observation with {\it Swift}/XRT started some time after the {\it
  Fermi} LAT trigger.  Using them, we derive a lower limit of the jet
break time to be $t_{X,j}<0.46$ days.
Providing that the afterglow emission in the submm  originated from
the same synchrotron radiation with the X-ray afterglow, the closure
relation requires the condition $\nu_{AMI}<\nu_{SMA}<\nu{a}$. Under
this condition, the radio afterglow should show decaying with
$\alpha=\sim-0.8$ and steeper spectral index of $\beta=2$. Although
the SMA light curve exhibited decaying, the brightening in the 15
GHz band with the corresponding temporal index of $\alpha=\sim0.33$ is
inconsistent with the relation. 
The radio spectral indices between the AMI and SMA bands
($\beta=\sim0.27$ at 2 days and $\beta=\sim0.05$ at 4 days) are too
flat and hence are inconsistent with the expected result.
Based on the closure relation, we also consider the two likely
conditions $\nu_{a}<\nu_{AMI}<\nu_{SMA}<\nu_{m}$ and
$\nu_{AMI}<\nu_{a}<\nu_{SMA}<\nu_{m}$ in the $p>2$ case.  The observed
results in the AMI (brightening) and SMA (steepness) bands are,
however, inconsistent with the temporal evolutions expected in either
of the conditions.
Therefore, we conclude that the radio emission originated from some
different radiation processes or regions from the X-ray emission.

\begin{figure*}[htb]
\centering
\includegraphics[width=1.8\columnwidth]{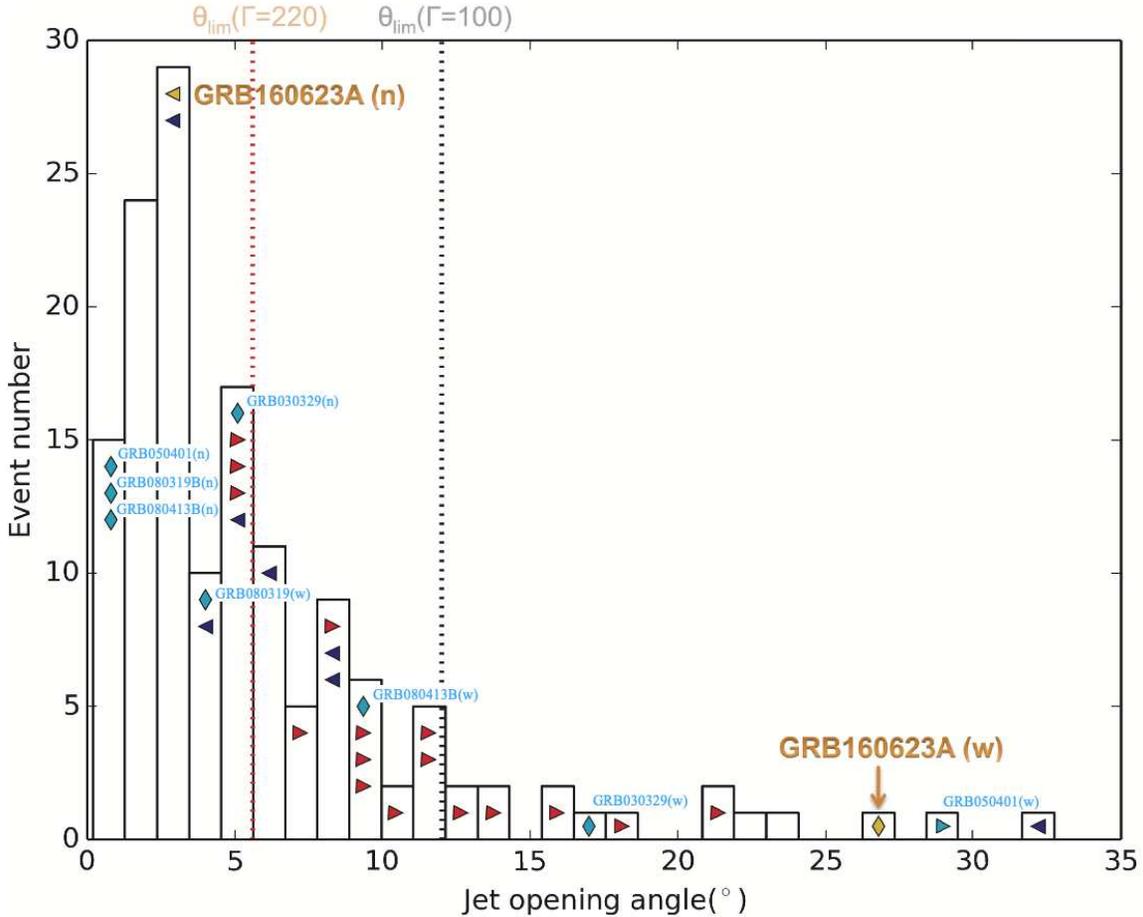}
\caption{Distribution of the jet opening angles  of GRB160623A and other GRBs. The narrow and wide jets of GRB160623A are highlighted with yellow marks. Upper and lower limits are indicated by blue and red arrows, respectively. The orange dotted line indicates the maximum jet opening angle for GRB160623A. The black dotted line indicates the maximum jet opening angle with $\Gamma_{0}$=100.
Four events reported as double jets (GRB030329, GRB050401, GRB080319B,
and GRB080413B) are also highlighted with cyan diamond marks.
We collected the jet opening angles of other GRBs from
literature \citep{racusin09,grb030329,frail01,bloom03,ghirlanda04,friedman05,cenko10,cenko11,filgas11}.
The measurement methods of their jet opening angles  are described in
individual references.  Basically, the methods  are identical to one another, as described in
\citet{sari99}  and according to afterglow modelling within the framework of
the forward-shock synchrotron radiation.
}
\label{jetangle}
\end{figure*}

We characterize the SMA and AMI light curves and spectra in the
forward-shock synchrotron-radiation framework. 
Since the optical light curve showed an unusual step decay
($\alpha_{opt}=\sim-4.6\pm0.3$) in the first day, we excluded the
optical data in the forward-shock modelling.
Employing the boxfit
code\citep{boxfit}, which is applicable in the on-axis configuration
with homogeneous circumburst medium (i.e. fixed observing angle as
$\theta_{obs}=0$), we obtain an optimal model with
$\theta_{jet}=27.7^{\circ}$, $E=7.7\times10^{52}$ erg, $n=70$ cm$^{\rm
  -3}$, $p$=2.6, $\epsilon_{B}=2.0\times 10^{-5}$, and
$\epsilon_{e}=1.9\times10^{-1}$.
These values are consistent with those of a typical GRB afterglow
\citep{panaitescu02,boxfit,huang17,xrf}, except for a wider jet opening
angle in our result than that of a typical GRB afterglow.
Note that the relatively higher circumburst density is consistent with
a high intrinsic absorption obtained from the X-ray spectrum of
GRB160623A \citep[e.g.][]{nh}. Figure \ref{jetangle} shows the
histogram of GRB jet opening angles. The jet opening angle of the
GRB160623A radio afterglow is largest among all GRBs.  Figure
\ref{lc-all} demonstrates that the model functions well describe the
observed radio light curves. With obtained physical parameters, we
also derive the expected light curves and confirm that the emission
from the wide jet in X-ray and optical bands should be negligible in
observations.


\subsection{Jet Opening angle and Cocoon radiation}

We further evaluate the jet opening angles on the basis of equation
(1) of \citet{frail01}, using the observed isolated equivalent energy
and assuming $\eta$ = 0.2, where $\eta$ is the radiative efficiency.
The jet opening angle for the radio afterglow is estimated, using the
temporal break in 230 GHz, to be
$\theta_{R,j}$=13$^\circ$.0$\pm$2$^\circ$.8 for the circumburst
density $n$ = 1 cm$^{-3}$ and 22$^\circ$.2$\pm$5$^\circ$.3 for $n$ =
70 cm$^{-3}$, where $n=1$ and $n=70$ cm$^{-3}$ are for the typical
value and for the estimated one from the radio afterglow modelling,
respectively.  Alternatively, using the explosion energy derived on
the basis of the afterglow modelling, we estimate the jet opening
angle of GRB160623A to be $26^{\circ}.3$.
These values are  more than twice larger than the typical jet opening
angle of the GRB. 
 In the same manner, we also estimate the upper limits of the jet opening angle for the X-ray
afterglow  to be $\theta_{X,j} < 2.8^{\circ}$ for $n$=1 cm$^{-3}$,
$\theta_{X,j} < 4.7^{\circ}$ for $n=70$ cm$^{-3}$, and  $\theta_{X,j}
< 5.6^{\circ}$ for $n=70$ cm$^{-3}$  from the explosion energy.
These upper limits are consistent with the typical value of GRB jet
opening angles (Figure \ref{jetangle}).

The origin of the wide jet emission may require an additional
component to those common for regular GRB afterglows.  \citet{anglelimit} constrained the maximum opening angle
$\theta_{j,max}$ to be 1/5$\Gamma_{0}$, where $\Gamma_{0}$ is the
initial Lorentz factor (i.e. $\theta_{j,max}<\sim$12$^{\circ}$ for
$\Gamma_{0}>100$).  We estimate the initial Lorentz factor of
GRB160623A to be $\Gamma_{0}>$ 220 from the prompt time variability of
0.250 sec\citep{lithwick01,golkhou15} and accordingly the maximum
opening angle of this event to be $\theta_{j,max}<5^{\circ}.5$.
In consequence, the radio afterglow jet angle of GRB160623A does not
agree with the theoretical maximum opening angle, whereas the upper
limit of the X-ray afterglow jet angle does.
According to \citet{nakar17}, the typical opening
angle of the relativistic cocoon afterglow is $\sim30^{\circ}$.  Since
the theory of the relativistic cocoon afterglow emission is similar to
that of the regular afterglow \citep{nakar17}, the parameters estimated
above characterize the shocked jet cocoon emission. 
Assuming the energy ratio of wide to narrow components as $E_{\rm
  wide}/E_{\rm narrow}\sim 0.1$ (i.e. the collapsar jet case;
\citet{peng}) and the identical micro-physical parameters ($n$,
$\epsilon_{B}$, and $\epsilon_{e}$) to the wide jet \citep{nakar17}
with the synchrotron slope of $p\sim2$ (based on the X-ray spectrum)
and the narrow jet opening angle of 5.5$^{\circ}$, we confirmed that
the expected narrow jet components in X-ray and radio bands can
describe the observed light curves (Figure \ref{lc-all}).
Considering the prompt phase of GRB160623A missed by {\it Fermi}/LAT
(Figure \ref{prompt-lc}), the huge total energy
($\sim8.5\times10^{53}$ erg) is likely reasonable as same as other
energitic ($>10^{54}$ erg) {\it Fermi}/LAT events
\citep[e.g.,][]{latgrb,latgrb2,latgrbcat}.  In fact, even the late
phase radiation in 100MeV--10GeV reached $(2.4\pm0.3)\times10^{52}$
erg \citep{latgrbcat}.  This result therefore implies that the
GRB160623A radio afterglow originated from a relativistic cocoon
afterglow.

The afterglows with double jet components are rarely observed. There
are only five events (shown in Figure \ref{jetangle}) and one of the
notable event is GRB030329 with $\theta_{n,j}=5^{\circ}.2$ and
$\theta_{w,j}=17^{\circ}.2$ \citep{grb030329}. Since afterglows of
GRB030329 and GRB160623A were densely monitored in the mm/submm
ranges, further mm/submm observations would address the wide jet and
shocked cocoon radiation.

\acknowledgments

This work is supported by the Ministry of Science and Technology of
Taiwan grants MOST 105-2112-M-008-013-MY3 (Y.U.).  This work made use
of data supplied by the UK Swift Science Data Centre at the University
of Leicester.

\facilities{Fermi, CALET, Swift, SMA, AMI}
\software{ }


\end{document}